# High-Performance Nanofluidic Osmotic Power Generation Enabled by Exterior Surface Charges under the Natural Salt Gradient

Long Ma,[1] Zhongwu Li,[2] Zhishan Yuan,[3] Haocheng Wang,[1] Chuanzhen Huang,[1] and Yinghua Qiu[1,4,5]*

1. Key Laboratory of High Efficiency and Clean Mechanical Manufacture of Ministry of Education, National Demonstration Center for Experimental Mechanical Engineering Education, School of Mechanical Engineering, Shandong University, Jinan, 250061, China

2. Jiangsu Key Laboratory for Design and Manufacture of Micro-Nano Biomedical Instruments, School of Mechanical Engineering, Southeast University, Nanjing 211189, China

3. School of Electro-mechanical Engineering, Guangdong University of Technology, Guangzhou, 510006, China

4. Advanced Medical Research Institute, Shandong University, Jinan, Shandong, 250012, China

5. Suzhou Research Institute, Shandong University, Suzhou, Jiangsu, 215123, China

*Corresponding author: yinghua.qiu@sdu.edu.cn

**Abstract:**

High-performance osmotic energy conversion (OEC) requires both high ionic selectivity and permeability in nanopores. Here, through systematical explorations of influences from individual charged nanopore surfaces on the performance of OEC, we find that the charged exterior surface on the low-concentration side (surface$_L$) is essential to achieve high-performance osmotic power generation, which can significantly improve the ionic selectivity and permeability simultaneously. Detailed investigation of ionic transport indicates that electric double layers near charged surfaces provide high-speed passages for counterions. The charged surface$_L$ enhances cation diffusion through enlarging the effective diffusive area, and inhibits anion transport by electrostatic repulsion. Different areas of charged exterior surfaces have been considered to mimic membranes with different porosities in practical applications. Through adjusting the width of the charged ring region on the surface$_L$, electric power in single nanopores increases from 0.3 to 3.4 pW with a plateau at the width of ~200 nm. The power density increases from 4200 to 4900 W/m$^2$ and then decreases monotonously that reaches the commercial benchmark at the charged width of ~480 nm. While, energy conversion efficiency can be promoted from 4% to 26%. Our results provide useful guide in the design of nanoporous membranes for high-performance osmotic energy harvesting.

**Keywords:** Osmotic Energy Harvesting, Natural Salt Gradient, Surface Charges, Electric Double Layers, Nanopores

# 1. Introduction

Sustainable development of our modern society demands abundant inexpensive clean energy. Osmotic energy also named "blue energy" provides us an important renewable energy resource, which is widely available from the salinity difference between two waterbodies, such as river water and seawater or salt lake water.[1] Due to the entropy change accompanying with mixing fresh and salty waters, osmotic energy is released, which could be harvested using porous membranes with ionic selectivity.[2] In practical applications, osmotic energy harvesting is constrained by the unsatisfied performance of permselective membranes in power density and energy conversion efficiency.[1]

In order to improve the performance of porous membranes for osmotic energy conversion (OEC), various porous membranes have been prepared, such as polymer membranes,[3, 4] anodic aluminum oxide membranes,[5] wood membranes,[6] multilayer membranes stacked with graphene oxide,[7] MXene,[8, 9] and boron nitride[10] nanosheets, as well as membranes formed with gel material,[11] and polymer networks.[12] The OEC study with porous membranes provides practical output power density, while the investigation with single nanopores sheds light on the OEC mechanism which could guide the design of nanoporous membranes because single nanopores are the elements in porous membranes. From theoretical predictions,[13] the induced power in OEC process is proportional to the total ionic flux and the transfer number of cations or anions, which are determined by the ion permeability and ionic selectivity of the nanopore, respectively. The OEC efficiency is controlled by ionic selectivity only.[13] In nanopores or nanochannels, ionic selectivity results from the electrostatic interactions between surface charges and mobile ions.[14] In order to reach high ionic selectivity, strong confinement, long length and dense surface charges of nanopores are usually required.[15] However, as the nanopore confinement and length increase, ion permeability decreases.[16]

Single nanopores in various materials have been fabricated. Based on the well-controlled fabrication process, track-etched polymer nanopores provide an important platform for early OEC study. Using a single conical polyimide nanopore, a maximum output power of 26 pW was achieved by Guo et al.,[2] when the load resistance was equal to the nanopore resistance. The output power also depended on the applied concentration gradients and solution pH, which determined the osmotic energy density and the surface charge density of pore walls, respectively. In order to promote the output power of single nanopores, many attempts have been conducted by increasing ionic selectivity, ion permeability, or both through various methods, such as chemical modification of pore walls,[17, 18] adjustment of the inner pore geometry,[19] and polymer adsorption to increase the space charge density.[12] As predicted by theories, ideal porous membranes for high-performance OEC should simultaneously possess high ion permeability and high ionic selectivity. With the creation of nanopores in low-dimensional materials, ultra-high power density had been predicted from single nanopores based on their strong surface charge density and ultra-thin thickness.[20-23]

In theoretical aspect, influences of nanopore properties on OEC could be investigated artificially. Surface charge density acts as an important parameter because it determines the ionic selectivity of the nanopore. As shown by finite element method (FEM), through increasing the surface charge density to 1 $C/m^2$, large output power could even be obtained in 100 nm-diameter nanopores due to the enhanced ionic selectivity.[24] Nanopore geometry plays a complex role because it affects both the ionic selectivity and ion permeability. In nanopores with a longer length or a smaller diameter, a higher ionic selectivity could be achieved, while in those shorter or wider ones, usually there is a larger ion flux.[25, 26] Through investigating the output power in nanopores with the same diameter but different lengths, optimal length range was found to induce the largest output power under specific conditions due to the balance between ionic selectivity and permeability of the nanopore.[13] Larger slip length of pore walls was also found to be able to improve the output power potentially, which lowered the viscous friction of fluid and enhanced the ion flux.[27, 28]

In the OEC study, simulations were mainly conducted using nanopores with above 200 nm in length, of which ionic selectivity was mainly determined by the charged inner surface.[13, 15, 29] The influence of charged exterior membrane surfaces has been rarely considered, which affect the ionic transport significantly in short nanopores, especially for the porous membranes with relative low porosities.[30-32] Here, taking advantages of FEM, the influences of individual charged surfaces in cylindrical nanopores on the OEC performance have been studied systematically. In order to guide practical applications, the natural salt gradient exists at the estuary of rivers was considered, i.e. seawater 0.5 M NaCl and river water 0.01 M NaCl. We found that the charged inner pore surface and exterior surface on the low-concentration side play the most important roles during the energy conversion. With the charged exterior surface on the low-concentration side, the output power could be improved over ~3.5 times than the maximum power in nanopores with only charged inner walls, with a efficiency as high as ~27%. Through detailed investigation of ionic flux distributions inside and outside the nanopore, electrical double layers provide a high-speed passage for the transport of counterions. The charged area in the exterior surface could promote the electric power significantly through simultaneously enhancing and inhibiting the diffusion processes of counterions and coions respectively. Based on our results, the performance in nanofluidic osmotic power generation can be improved effectively through suitable methods of surface modification.

**2. Simulation Methods**

COMSOL 5.2 package was used to solve our 3D simulations, in which coupled Poisson-Nernst-Planck (PNP) and Navier-stokes (NS) equations were taken into account.[33, 34] As shown in Figure 1, two reservoirs with 5 μm in radius and 5 μm in length are connected to both sides of the nanopore. Simulations were conducted at room temperature of 298 K. The relative permittivity of water was set as 80. In the simulations, pore lengths varying from nanoscale to microscale were considered, which ranged from 1 to 1000 nm. The pore radius was set as 5 nm which could be conveniently fabricated in experimental reports.[3, 5, 35-37] Due to the much smaller sizes of ions[38] than the pore diameter, the effect of ion sizes on the ionic transport was not considered. In order to correlate simulations with practical applications, the natural salt gradient exists at the estuary was used in our simulations, i.e. NaCl solutions in 500 mM and 10 mM.[9, 39] The surface charge density was considered as −0.08 C/m$^2$.[13, 30, 40] 1.33×10$^{-9}$ and 2.03×10$^{-9}$ m$^2$s$^{-1}$ were used for the diffusion coefficients of Na$^+$ and Cl$^-$ ions ($D_{Na^+}$ and $D_{Cl^-}$).[41] As shown in Figure 1, under the concentration gradient, three surfaces of nanopores i.e. inner surface and exterior surfaces on high- and low-concentration sides are denoted as surface$_{inner}$, surface$_H$ and surface$_L$, respectively. Based on the charged statuses, w/ and w/o surface charges on the three individual surfaces, 8 different simulation models were considered, i.e. nanopores with no charged surfaces (NCS), charged surface$_{inner}$ (ICS), charged surface$_H$ (ECS$_H$), charged surface$_L$ (ECS$_L$), charged surface$_{inner}$ and surface$_H$ (IECS$_H$), charged surface$_{inner}$ and surface$_L$ (IECS$_L$), charged exterior surfaces (ECS), as well as uniformly charged surfaces (ACS). The subscript H or L in our definition represents the high or low concentration.

Table S1 lists corresponding boundary conditions for the 8 models. In order to consider the effect of electric double layers, 0.1 nm mesh size was used for the inner pore surface and 3 μm-wide regions of both exterior surfaces near the pore boundary. For other parts in exterior surfaces, the mesh size of 0.5 nm was chosen to lower the memory cost. The strategy of mesh building is shown in Figure S1. The total mesh nodes used in the simulations were above 1,000,000. Through solving the same problem studied by Cao et al.,[13] our simulation model had been verified. As plotted in Figure S2, our results agree well with theirs.

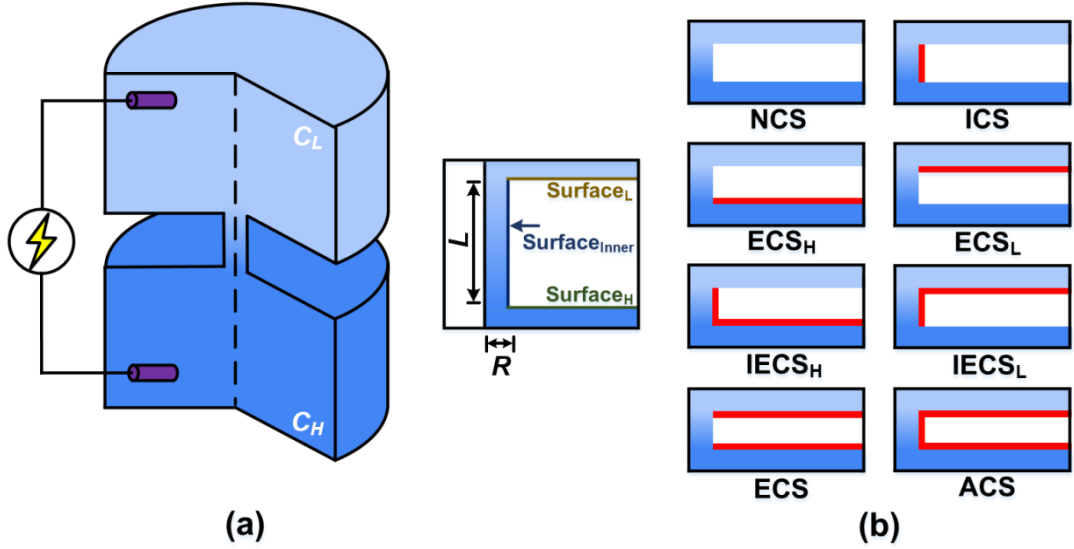

(a)    (b)

Figure 1 (a) Simulation scheme under the concentration gradient. $C_H$ and $C_L$ denote the high and low concentrations, respectively. Both reservoirs are 5 μm in radius and 5 μm in height. Inset shows the zoomed-in cylindrical nanopore region. $L$ and $R$ represent the length and radius of nanopores. Three surfaces of the nanopore are denoted as surface$_{inner}$, surface$_H$ and surface$_L$. (b) 8 simulation models with different charged surfaces. Surface charges considered in this work are always negative which are shown in red.

In the simulations of the OEC process, induced membrane potential ($V_0$)[13] was calculated with eq. 1. As shown in Figure S3, theoretical prediction of $V_0$ from eq. 1 has almost the same value as that obtained from current-voltage curves which is the typical method used in experiments.[2, 12]

$$V_0 = (2t_+ - 1)\frac{RT}{F}\ln\frac{\alpha_H}{\alpha_L} \qquad (1)$$

where *R*, *T*, and *F* are the gas constant, temperature and Faraday constant, respectively. $α_H$ and $α_L$ denote the chemical activity of the ions on both high and low concentration sides. Salt concentrations of 500 and 10 mM were used for $α_H$ and $α_L$. $t_+$ is the cation transfer number,[15] i.e. the ionic selectivity to cations, which is obtained with eq. 2.

$$t_+ = |I_+|/(|I_+|+|I_-|) \tag{2}$$

in which $|I_+|$ and $|I_-|$ are the current values contributed by cations and anions respectively. The electric power ($P_{max}$)[2, 42] was predicted from eq. 3.

$$P_{max} = \frac{1}{4}I_0 V_0 \tag{3}$$

Steady-state diffusion current ($I_0$) was integrated from the boundary of the reservoir. In this work, diffusion current was defined as the total ionic current through the nanopores under concentration gradients. The OEC efficiency [13] was calculated with eq. 4.

$$\eta = (2t_+ - 1)^2 \tag{4}$$

**3. Results and Discussions**

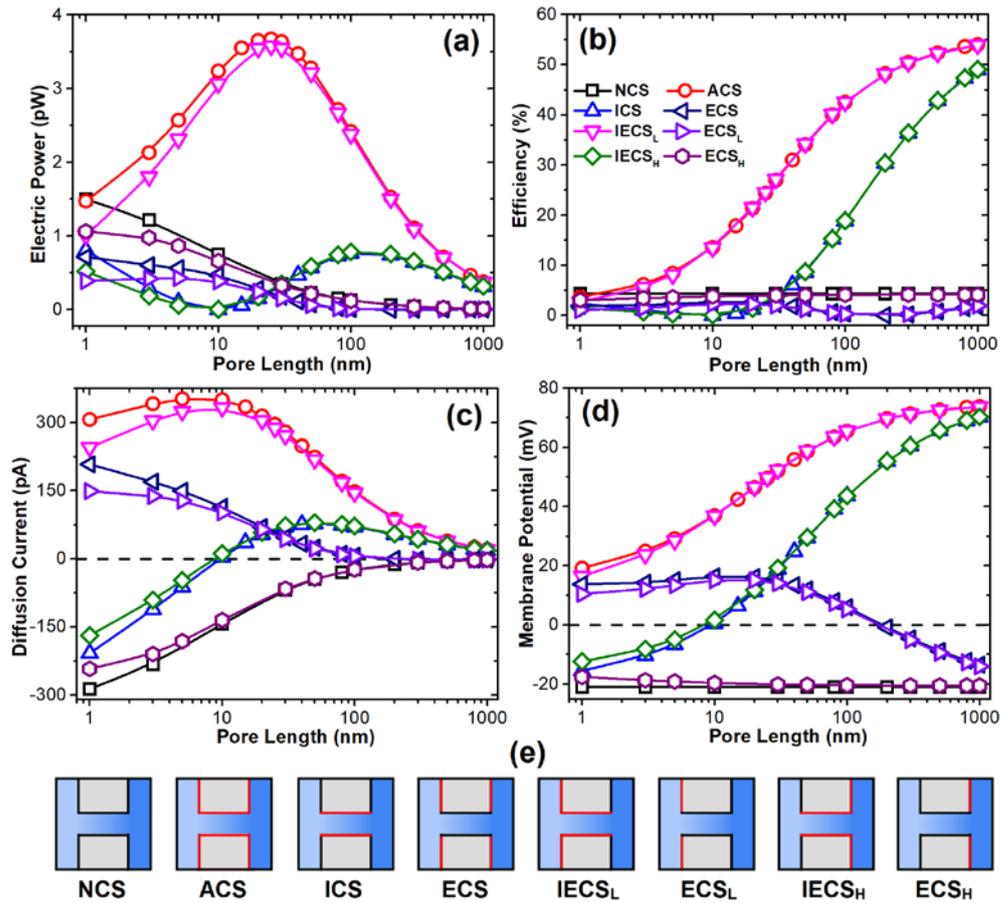

Figure 2 Performance of osmotic energy conversion in single nanopores from 8 different simulation models with various pore lengths. (a) Electric power, (b) OEC efficiency, (c) diffusion current, and (d) membrane potential. (e) Schemes of simulation models with different charged surfaces shown in red. Pore radius was set as 5 nm. Solutions on both sides of nanopores were 0.5 M and 0.01 M NaCl. Surface charge density was −0.08 C/m$^2$.

Figure 2 shows the OEC performance of nanopores with different charged surfaces and various lengths. When no surface charges appear on pore walls (NCS model), diffusion current is dominated by Cl$^-$ ions due to its larger diffusion coefficient[41] which results in a negative membrane potential of ~−20 mV.

Consequently, electric power could be generated but with a low efficiency of ~4%. With the pore length increasing, the output power decreases because of the inhibited diffusion flux of ions across the nanopore. While the stable membrane potential results from the constant cation selectivity of ~40% in the nanopore, which equals to $D_{Na^+}/(D_{Na^+}+D_{Cl^-})$.

When the nanopore gets negatively charged on one or more surfaces, Na$^+$ and Cl$^-$ ions are attracted or respelled by the surface charges, respectively. Because of its high strength and long range,[43] electrostatic interactions between surface charges and mobile ions play important roles in ionic transport. Following, the influences of three individual charged surfaces on the OEC performance have been investigated.

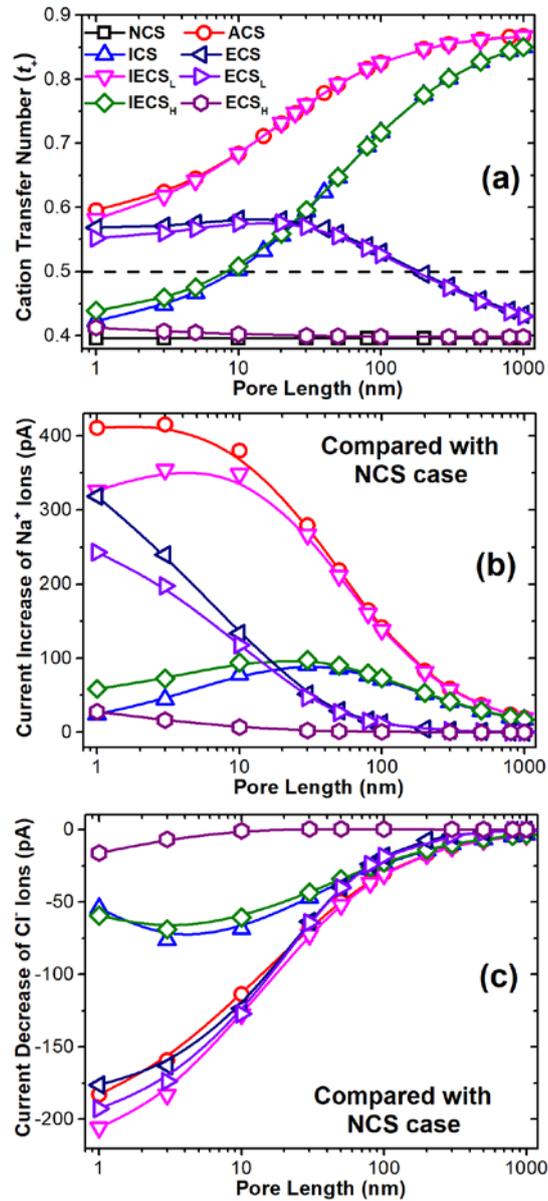

Figure 3 Ionic transport characteristics in single nanopores from 8 different simulation models with various pore lengths. (a) Cation transfer number $t_+$, (b) increase of $Na^+$ ions current, and (c) decrease of $Cl^-$ ions current in models with one or more charged surfaces than ionic current from the NCS model.

As shown in Figure 2c, in the case with only charged inner surface (ICS model), when the pore length is less than ~10 nm, diffusion current is mainly contributed by Cl$^-$ ions due to the ignored cation selectivity (Figure 3a) which also induces a negative membrane potential. As the pore length increases, larger charged inner surface area provides stronger electrostatic interactions to the mobile ions, which promotes cation selectivity significantly.[15] The diffusion current and membrane potential consequently vary from negative values to 0, then increase to positive values with the pore length expanding further. These result in a decrease-increase profile of the electric power. After the nanopore length exceeds ~100 nm, diffusion current starts to reduce due to the inhibited ionic flux, which leads to the decrease of electric power. Based on the monotonic enhanced cation transfer number ($t_+$) from ~0.4 to ~0.85, the OEC efficiency first drops from ~4% to 0 and then increases to ~50%. The increase-decrease trend of output power during the variation of pore length from ~10 to 1000 nm agrees well with the simulation results of Cao et al.[13] However, in the earlier report, the power drop with the pore length varying from 1 to ~10 nm was not observed because in their work negatively charged nanopores were always cation-selective in KCl solutions.

Simulation reports on the OEC or ionic selectivity were mainly conducted with long nanopores, in which exterior surface charges were usually ignored[2, 15, 17] due to the dominated contribution of charged inner surfaces to the ionic selectivity. Under the concentration gradients across nanopores, here, two different simulation models were used to consider each charged exterior surface separately, i.e. nanopores with charged exterior surface on the high-concentration side (ECS$_H$ model) or low-concentration side (ECS$_L$ model). From Figure 2 and 3, through comparison of the diffusion current and membrane potential obtained from $ECS_H$ and NCS cases, the charged surface$_H$ has little influence on the ionic transport or the OEC performance. However, when the nanopore is extremely thin, such as 1 nm, the output power shows a small drop, which is caused by the reduced diffusion current and membrane potential in small magnitudes.

However, in the $ECS_L$ case the charged surface$_L$ has significant influence on the diffusion current and membrane potential. As shown in Figure 2 and 3, it enables the nanopore cation selectivity when the nanopore is shorter than ~200 nm. With the pore length increasing to 1000 nm, its influence gets suppressed gradually. In the range of the pore length from 1 to ~30 nm, the charged surface$_L$ can promote the cation transfer number by 15% than the NCS case. Since the increased cation transfer number is closer to 50%, it results in a reduced OEC efficiency. Based on the small values of the diffusion current and membrane potential, the output power is lower than that in the NCS model.[13]

Under concentration gradients across the nanopore, the weak influence by charged surface$_H$ could be caused by the strong screening of the counterions.[14] While in the simulations under a lower salt gradient of 0.05 M/0.01 M NaCl solutions (Figure S4), the charged surface$_H$ still has much weaker promotion in OEC performance than that caused by the charged surface$_L$, due to the weakly enhanced ionic flux and cation selectivity of the nanopore. Since the charged surface$_H$ cannot promote the transport of cations as the charged surface$_L$ does (as shown below), its influence is limited and can be neglected when the nanopore is longer than ~50 nm.

With combinations of individual charged surfaces, we studied the OEC performance in nanopores with two or all charged surfaces. Due to the neglected influence of the charged surface$_H$ on the diffusion current and membrane potential, in the cases with charged surface$_{inner}$ and surface$_H$ (IECS$_H$ model) or both charged exterior surfaces (ECS model), the ion transport and OEC performance were determined by the charged inner pore surface and exterior surface on the low-concentration side, respectively.

Near charged surfaces, electric double layers (EDLs) forms spontaneously where counterions and coions are accumulated and depleted, respectively. In nanopores with negatively charged inner walls, cation flux contributed from EDLs results in the cation selectivity, especially in long nanopores.[15] While in those shorter ones, due to the relatively large ion permeability, besides charged inner surfaces, charged exterior surfaces are essential to achieve high cation selectivity.[30, 31] From Figure 3a, when the pore is longer or shorter than ~25 nm, the cation selectivity of nanopores is mainly determined by the charged surface$_{inner}$ or surface$_L$, respectively.[15, 31] When both charged surfaces were considered together (IECS$_L$ model), the diffusion current contributed by Na$^+$ ions and Cl$^-$ ions was significantly promoted and suppressed simultaneously, which induced a much larger cation selectivity. Accompanying with the increase-decrease variation of diffusion current with the pore length, the output power shows an increase-decrease trend with its maximum appearing at ~25 nm in length, which reaches ~4.6 times of the maximum power in the ICS case. Due to the gradually increased cation selectivity from ~0.6 to ~0.9 with pore length varying from 1 to 1000 nm, the efficiency increase monotonously from ~4% to above 50%. When the pore is uniformly charged (ACS model), its OEC performance is very similar to that of IECS$_L$ model because of the ignored effect of the charged surface$_H$.

From our simulations, under the natural salt gradient, the IECS$_L$ and ACS models provide the best OEC performance, in which the charged surface$_{inner}$ and surface$_L$ are of great importance. This OEC performance shown here does not depend on the fluid flow in the nanopore. In the simulations using the ACS model without the consideration of NS equations (Figure S5), the electric power and efficiency show almost the same trend as those obtained in the cases including NS equations.

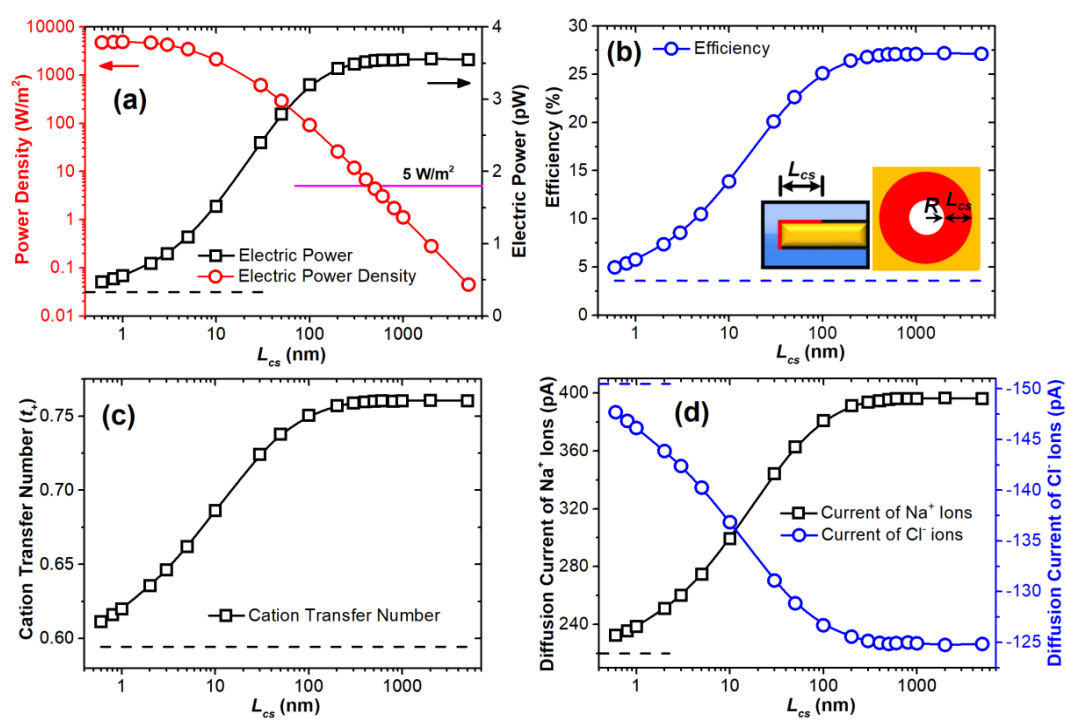

Figure 4 Performance of the OEC and characteristics of ionic transport in single nanopores with charged inner surface and exterior surface on the low-concentration side (IECS$_L$ model) with various widths ($L_{cs}$) of charged ring regions on the surface$_L$.

(a) Electric power and electric power density. Power density was calculated through $P_{max}/[\pi(R + L_{cs})^2]$. Pink line shows the power density of 5 W/m$^2$ which is the commercialization benchmark. (b) OEC efficiency. Inset shows the simulation scheme. Surface charges are shown in red. (c) Cation transfer number ($t_+$). (d) Diffusion current contributed from Na$^+$ and Cl$^-$ ions. Dashed lines in each figure shows the values obtained from the ICS model. Pore radius and length were 5 and 30 nm, respectively.

Based on the significant enhancement of electric power induced by the charged surface$_L$, we conducted a series of simulations to investigate the dependence of the OEC performance on the charged area around the orifice with the IECS$_L$ model.[30, 31] It is of great importance in practical applications, because in membranes with a high pore density multipores behave in a parallel way. The OEC performance of individual nanopores could be influenced by nearby nanopores.[40, 44] As shown in Figure 4, for nanopores with 5 nm in radius and 30 nm in length, with the charged width ($L_{cs}$) increasing from 0 to ~200 nm, diffusion current of cations and anions gets enhanced and suppressed by ~77% and ~17%, which promotes the cation transfer number from ~0.59 to ~0.76. Consequently, the output power and energy conversion efficiency of single nanopores are improved by ~9.3 and ~6.4 times, respectively. For nanopores with fully charged exterior surfaces considered in Figure 2 and 3, the effective charged width needs to be ~200 nm to reach their best performance. This finding agrees with an earlier simulation report with ultra-thin nanopores.[30] In the control simulations with the ECS$_L$ model (Figure S6), almost the same effective charged width was found.

In order to extend our simulation results of single nanopores to the OEC performance during practical applications using porous membranes, the power density was calculated through $P_{max}/[\pi(R+L_{cs})^2]$ under a hypothesis that nanopores in porous membranes were neatly arranged.[40] With $L_{cs}$ increasing from 0 to ~1 nm, the power density has a small enhancement (Figure S7, the power density is shown in the linear plot). While as the charged width expands further to 5 μm, the power density decreases monotonously, due to the faster growth in area than that of electric power. After $L_{cs}$ becomes larger than ~400 nm, the power density gets below the commercial benchmark of 5 W/m$^2$.[1] However, the OEC efficiency is enhanced from 4% to ~27.5%.

Ionic selectivity caused by charged inner pore surfaces has been explored extensively, which could affect the concentration distributions of ions inside nanopores. While, the mechanism that charged exterior surfaces control ionic selectivity of thin nanopores has been rarely explored. From our simulation models, charged exterior surfaces provide weak electrostatic interaction to mobile ions directly during their passage through the nanopores because most of the surface area locates far away from orifice. In addition, near charged surfaces the range of electrostatic interaction is usually within several nanometers.[43] Our finding shown in Figure 4 is unexpected because the charged exterior surface 100 nm away from the nanopore opening could significantly influence the ionic transport in the nanopores.

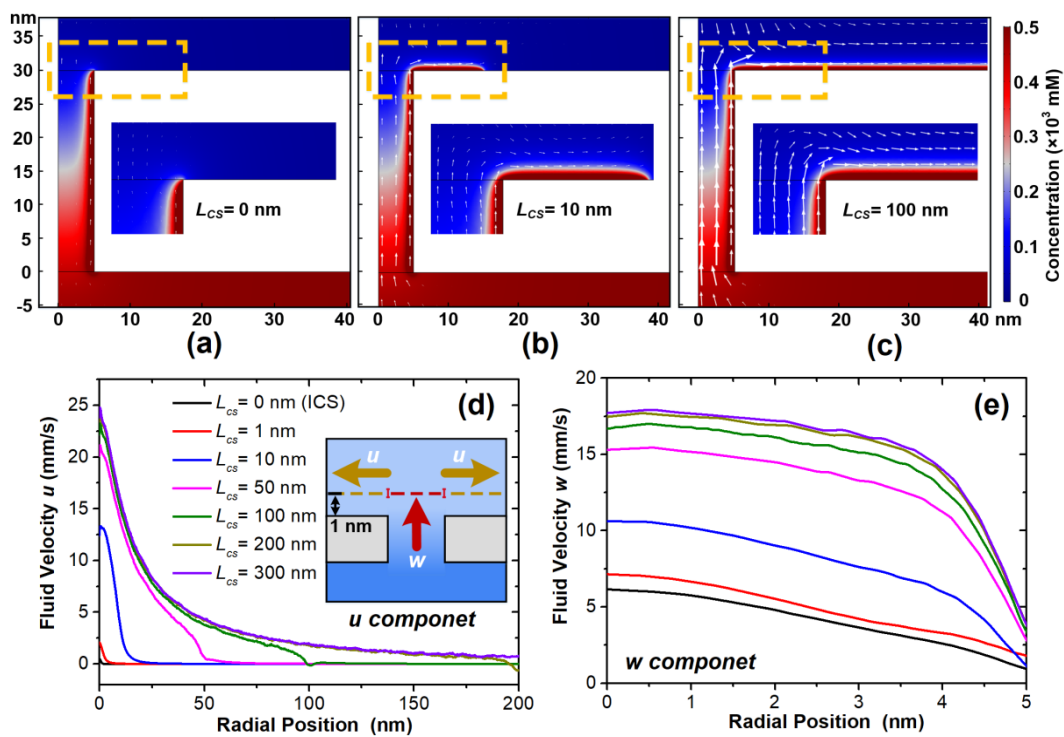

Figure 5 (a-c) 2D concentration distributions of Na$^+$ ions (color maps) and fluid flow (white arrows, all in the same scale) in the pore and near the exterior membrane surface on the low-concentration side from 3 cases with different $L_{cs}$ from the IECS$_L$ model. (a) $L_{cs}$=0 nm, i.e. the ICS model. (b) $L_{cs}$=10 nm, and (c) $L_{cs}$=100 nm. (d) Fluid velocity (radial component $u$) distributions along the exterior membrane surface with various $L_{cs}$. Inset shows the scheme of two fluid flow components: axial component $w$, and radial component $u$. (e) Fluid velocity (axial component $w$) distributions in the radial direction above the pore region under various $L_{cs}$. Velocity distributions in (d) and (e) were obtained in the plane of 1 nm above the membrane surface as shown in the scheme.

From Figure 4c, nanopores are cation selective with various charged widths in the exterior surface. Taking advantages of the NS equations considered in our simulations, fluid flow i.e. diffusio-osmotic flow[45] induced under concentration gradients could be plotted in Figure 5, which can shed light on the transport of counterions in the nanopore and near the charged exterior surfaces. Here, the two components of fluid flow are defined as the axial component $w$, and radial component $u$. When no surface charges appear on the exterior surface (ICS model), the fluid flow mainly happens inside the nanopore which is along the axis and very weak (Figure 5d and 5e). After the surface$_L$ starts to get charged, the fluid starts to flow parallel to the exterior surface whose magnitude enhances with $L_{cs}$. Simultaneously, the axial flow velocity gets promoted. After $L_{cs}$ expands to ~200 nm, the velocity of fluid flow in both axial and radial directions achieves their saturation.

Figure S8 exhibits the ion concentration distributions in the plane of 1 nm above the membrane surface which locates inside the EDLs (Scheme shown in Figure 5d). Near negatively charged surfaces, Na$^+$ and Cl$^-$ ions get accumulated and depleted. After $L_{cs}$ is larger than 10 nm, the concentration of Na$^+$ and Cl$^-$ ions increases to ~180 mM and reduces to ~0 mM, respectively. Due to the strong fluid flow parallel to the membrane surface, the radial ionic fluxes contributed by Na$^+$ and Cl$^-$ ions are enhanced and suppressed. As the charged width reaches ~200 nm, the speed of fluid flow becomes saturated, which accompanies the largest cation current and the lowest anion current, as depicted in Figure 4d. Consequently, the electric power and efficiency are determined, which reach ~96% and ~97% of those with fully charged exterior surface, respectively. Relative OEC performance from cases with various $L_{cs}$ to the one with $L_{cs}$ =5 μm is shown in Figure S9. From our control simulations, the opposite charged pattern to that shown in the inset of Figure 4b have been considered, i.e. a neutral ring area with its width ($L_{ns}$) near the pore opening but the rest of the exterior membrane is charged (Figure S10). With $L_{ns}$ increasing to 10 nm, the electric power decreases to the value obtained in the ICS case. This is due to that the promotion in the transport of counterions along the charged exterior surface is cut off.

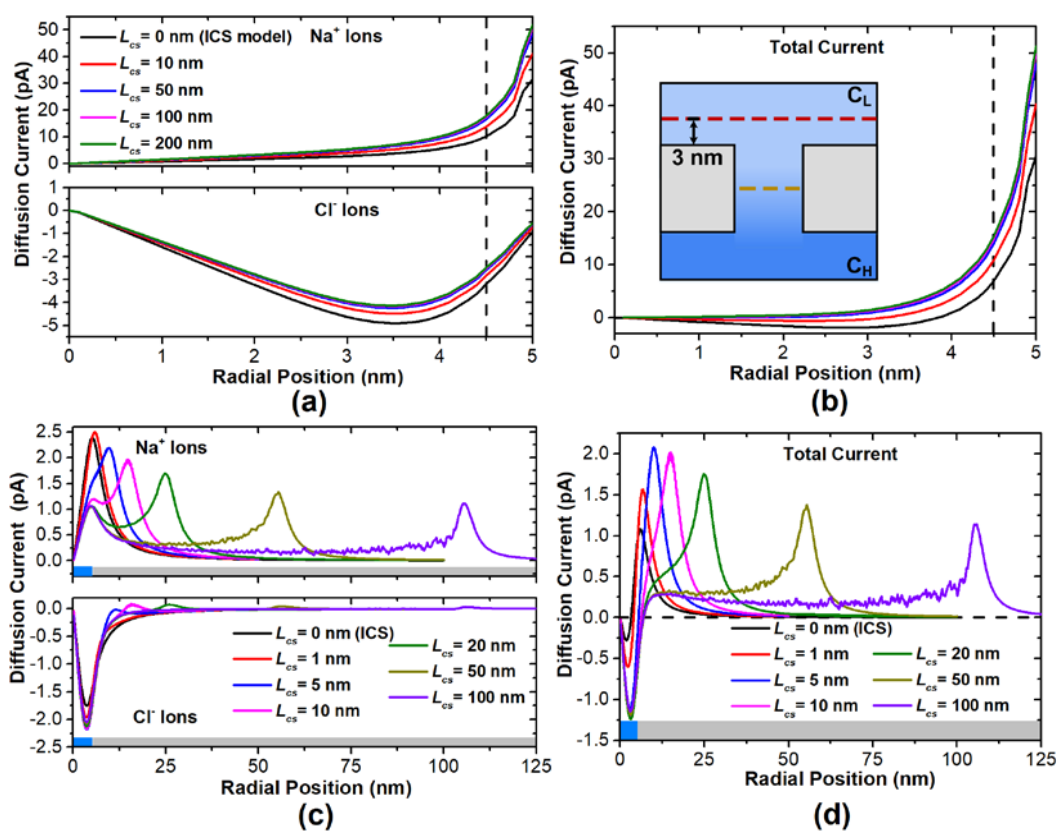

Figure 6 Diffusion current distributions inside and outside the nanopore with various widths ($L_{cs}$) of charged ring regions on the exterior membrane surface at the low-concentration side (IECS$_L$ model). (a-b) Distributions of diffusion current in the center cross-section of the nanopores. (a) Na$^+$ and Cl$^-$ ions current, and (b) total current. (c-d) Distributions of diffusion current in the plane of 3 nm above the membrane surface, i.e. the boundary of EDLs region. (c) Na$^+$ and Cl$^-$ ions current, as well as (d) total current. Red and brown lines in the inset of (b) show the locations where current distributions were obtained. Light blue and grey regions at the bottom of each panel in (c) and (d) represent the positions of the nanopore and membrane.

In order to investigate the ionic transport more quantitatively, the ionic current distributions along the radial direction have been explored in the center cross-section of the pore and the boundary surface of the EDLs i.e. the plane of 3 nm above the exterior membrane surface (Scheme shown in Figure 6). Each point was obtained through integration of the ionic flux over 0.1-nm-wide segments. As plotted in Figure 6a, due to the negatively charged pore walls, cations act as the main current carriers, most of which transport within 0.5 nm away from the pore surface.[24] While $Cl^-$ ions as the coions are repelled from surface charges, passing the nanopore mainly in the ring-region of 0.5–3 nm above the pore wall. The flux distributions inside the nanopore agree well with the ion distributions in EDLs.[14] As $L_{cs}$ expands from 0 to 200 nm, total ionic current increases due to the flux enhancement of $Na^+$ ions.

Based on the continuity of ionic current through the nanopore, all ions flowing through the center pore cross-section will pass the exit and enter the reservoir. From Figure 6c, in the ICS case, the flux of $Na^+$ ions crossing the plane of 3 nm above the membrane surface mainly locates at ~5 nm in the radial direction, i.e. the position of the inner pore wall. When the exterior surface gets negatively charged, counterions start to flow parallel to the membrane surface, which splits the peak of $Na^+$ ions flux near the pore opening into two individual smaller ones: one keeps at the same location, the other one appears near the end of the charged region. Above the vast charged exterior surface, diffusion flux of cations also exists due to the concentration gradient between the EDLs and bulk regions. While the transport of Cl⁻ ions from the nanopore into the reservoir mainly happens within 5 nm away from the pore axis, which is weakly enhanced by exterior surface charges. From distributions of total ionic current in Figure 6d, due to the balance off between currents contributed by cations and anions, within 5 nm away from the axis Cl⁻ ions dominate the net ionic flux. It's surprising to find the effective diffusion current for the OEC occurring above the membrane surface, which has a distinct peak locating at the farther boundary of the charged region.

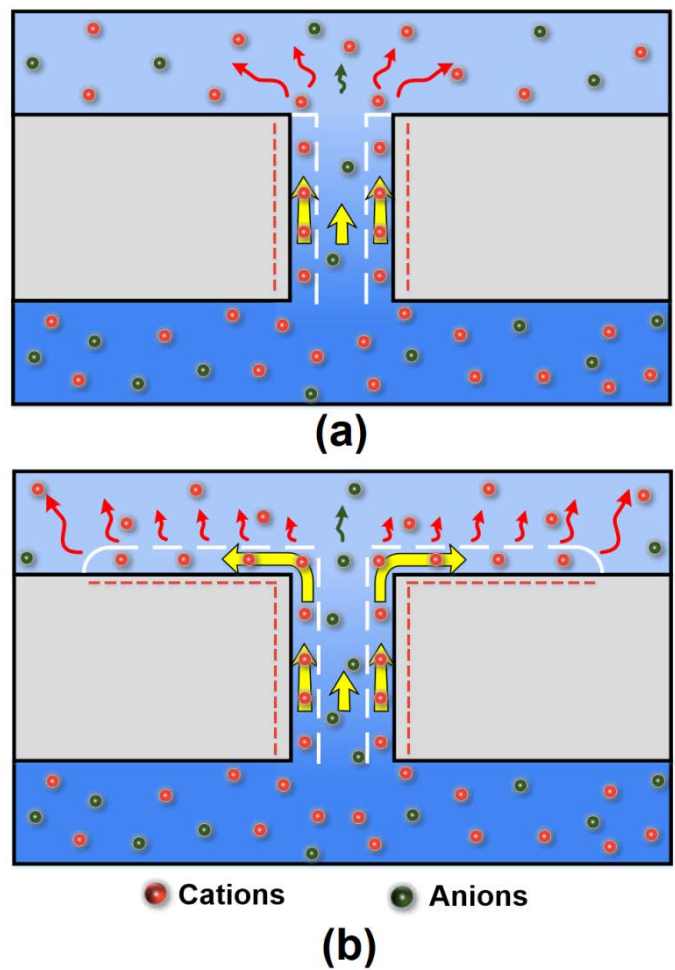

Figure 7 Illustrations of the ionic transport during OEC under salt concentration gradients. (a) ICS model, and (b) IECS$_L$ model. Yellow arrows shows the diffusio-osmotic flow caused by the cation movement. Curved orange and green arrows represent the ionic diffusion of cations and anions. Dashed red lines indicate the surface charges.

The mechanisms of ion transport governed by the charged inner pore surface and exterior surface on the low-concentration side are concluded in Figure 7. Due to appearance of surface charges, EDLs regions provide an effective passage for the transport of concentrated counterions.[24] Under the concentration gradient across the nanopore (Figure 7), cations are motivated to form the ionic flow along charged surfaces.[45] Coions in small amount translocate the nanopore in the center region. When the exterior membrane becomes charged, due to the electrostatic attraction, counterions start to flow parallel to the membrane surface,[46] which could diffuse away from the EDLs region to the bulk due to the concentration gradients, especially at the end of the charged surface. From the potential distribution along the exterior surface, the electric potential becomes more negative in the radial direction (Figure S11). This is due to the weaker electrostatic screening caused by the faster diffusion of Na$^+$ ions away from the EDLs in the area farther away from the orifice. Since the surface area for ionic diffusion enlarges with the charged width expanding, total diffusion current increases. Control simulations without the consideration of NS equations were also conducted, with $L_{cs}$ varies from 0 to ~200 nm, electric power and OEC efficiency share almost the same trend as those in the cases considering NS equations (Figure S12) which means that the enhancement in cation diffusion process is not determined by the convective flux caused by fluid flow.

However, with $L_{cs}$ approaching ~200 nm, the total diffusion process achieved a plateau. We think this is attributed by the limited ionic flux inside the nanopore, because the charged exterior membrane possesses enough area for more ions to diffuse into the reservoir. Due to the almost unchanged concentration gradient across the nanopore with various $L_{cs}$, the ionic flux inside the nanopore saturates at $L_{cs}$ = ~200 nm (Figure S13). This may be proved by experiments with ultrathin nanopores or nanopores with slipping inner surfaces. For single $MoS_2$ nanopores, the effective charged area on the membrane surface around the orifice could reach up to micrometers which provides significant improvement to the cation diffusion.[30] The obtained ultra-high power is caused by the relative larger ionic flux inside the nanopore due to the ultrathin membrane thickness.[21] When the inner pore surface becomes slipping, electric power could also be promoted by the enhanced ionic flux due to the weak viscous force between fluid flow and pore surface.[27, 28]

## 4. Conclusions

From systematical investigation with FEM, the effects of individual charged nanopore surfaces on the OEC process have been investigated under the natural salt gradient. We found that the charged $surface_{inner}$ and $surface_L$ play important roles in the high-performance osmotic energy generation. When the nanopore is longer than ~25 nm, the ionic transport is governed by the charged $surface_{inner}$. While in shorter pores, the charged surface$_L$ dominates the ionic selectivity. Ultra-high electric power could be achieved in the nanopores with both charged surface$_{inner}$ and surface$_L$. Based on the tradeoff between the ionic selectivity and ion permeability, the optimal pore length for the OEC varies from ~20 to ~30 nm, which provides the output power above ~4.5 times of the maximum power generated in the case with only charged surface$_{inner}$. Through analyzing the flux distribution inside and outside the nanopore, EDLs regions near both charged surfaces provide a high-speed passage for the transport of counterions. Due to the considerable promotion and inhibition in cation and anion diffusion processes respectively, the charged area on the exterior membrane surface could improve the electric power and energy conversion efficiency of single nanopores significantly. Through taking consideration of widths of charged exterior surface into the effective pore area, our simulation results could be correlated to the practical performance of membranes with various porosities based on the parallel behavior of multipores. Though as the width of the charged ring area expands from ~1 nm to ~480 nm the power density decreases monotonously, it is still larger than the commercial benchmark and the energy conversion efficiency could be promoted from 4% to 26%. Based on the detailed mechanism of ionic transport governed by the charged surfaces, we think our results could shed light on the design of porous membranes for high-performance osmotic energy harvesting.

**Supporting Information:**

Simulation details, model verification, membrane potential and diffusion current, and more supported simulation results.

**Acknowledgments:**

This research was supported by the Shandong Province Natural Science Foundation (ZR2020QE188), the Natural Science Foundation of Jiangsu Province (BK20200234), the Guangdong Basic and Applied Basic Research Foundation (2019A1515110478), the Qilu Talented Young Scholar Program of Shandong University, Key Laboratory of High-efficiency and Clean Mechanical Manufacture at Shandong University, Ministry of Education, and the Open Foundation of Advanced Medical Research Institute of Shandong University (Grant No. 22480082038411).

**References:**

[1] A. Siria, M.-L. Bocquet, L. Bocquet, New Avenues for the Large-Scale Harvesting of Blue Energy. Nat. Rev. Chem. 1 (2017) 0091. https://doi.org/10.1038/s41570-017-0091.
[2] W. Guo, L. Cao, J. Xia, F. Q. Nie, W. Ma, J. Xue, Y. Song, D. Zhu, Y. Wang, L. Jiang, Energy Harvesting with Single-Ion-Selective Nanopores: A Concentration-Gradient-Driven Nanofluidic Power Source. Adv. Funct. Mater. 20 (2010) 1339-1344. https://doi.org/10.1002/adfm.200902312.
[3] Z. Zhang, X.-Y. Kong, K. Xiao, Q. Liu, G. Xie, P. Li, J. Ma, Y. Tian, L. Wen, L. Jiang, Engineered Asymmetric Heterogeneous Membrane: A Concentration-Gradient-Driven Energy Harvesting Device. J. Am. Chem. Soc. 137 (2015) 14765-14772. https://doi.org/10.1021/jacs.5b09918.
[4] K. Kwon, S. J. Lee, L. Li, C. Han, D. Kim, Energy Harvesting System Using Reverse Electrodialysis with Nanoporous Polycarbonate Track-Etch Membranes. Int. J. Energy Res. 38 (2014) 530-537. https://doi.org/10.1002/er.3111.
[5] J. Kim, S. J. Kim, D.-K. Kim, Energy Harvesting from Salinity Gradient by Reverse Electrodialysis with Anodic Alumina Nanopores. Energy 51 (2013) 413-421. https://doi.org/10.1016/j.energy.2013.01.019.
[6] Q.-Y. Wu, C. Wang, R. Wang, C. Chen, J. Gao, J. Dai, D. Liu, Z. Lin, L. Hu, Salinity-Gradient Power Generation with Ionized Wood Membranes. Adv. Energy Mater. 10 (2020) 1902590. https://doi.org/10.1002/aenm.201902590.
[7] J. Ji, Q. Kang, Y. Zhou, Y. Feng, X. Chen, J. Yuan, W. Guo, Y. Wei, L. Jiang, Osmotic Power Generation with Positively and Negatively Charged 2d Nanofluidic Membrane Pairs. Adv. Funct. Mater. 27 (2017) 1603623. http://doi.org/10.1002/adfm.201603623.
[8] L. Ding, D. Xiao, Z. Lu, J. Deng, Y. Wei, J. Caro, H. Wang, Oppositely Charged Ti3c2tx Mxene Membranes with 2d Nanofluidic Channels for Osmotic Energy Harvesting. Angew. Chem. Int. Ed. 59 (2020) 8720-8726. https://doi.org/10.1002/anie.201915993.

[9] Z. Zhang, S. Yang, P. Zhang, J. Zhang, G. Chen, X. Feng, Mechanically Strong Mxene/Kevlar Nanofiber Composite Membranes as High-Performance Nanofluidic Osmotic Power Generators. Nat. Commun. 10 (2019) 2920. https://doi.org/10.1038/s41467-019-10885-8.

[10] C. Chen, D. Liu, L. He, S. Qin, J. Wang, J. M. Razal, N. A. Kotov, W. Lei, Bio-Inspired Nanocomposite Membranes for Osmotic Energy Harvesting. Joule 4 (2020) 247-261. https://doi.org/10.1016/j.joule.2019.11.010.

[11] Z. Zhang, L. He, C. Zhu, Y. Qian, L. Wen, L. Jiang, Improved Osmotic Energy Conversion in Heterogeneous Membrane Boosted by Three-Dimensional Hydrogel Interface. Nat. Commun. 11 (2020) 875. https://doi.org/10.1038/s41467-020-14674-6.

[12] C.-Y. Lin, C. Combs, Y.-S. Su, L.-H. Yeh, Z. S. Siwy, Rectification of Concentration Polarization in Mesopores Leads to High Conductance Ionic Diodes and High Performance Osmotic Power. J. Am. Chem. Soc. 141 (2019) 3691-3698. https://doi.org/10.1021/jacs.8b13497.

[13] L. Cao, F. Xiao, Y. Feng, W. Zhu, W. Geng, J. Yang, X. Zhang, N. Li, W. Guo, L. Jiang, Anomalous Channel-Length Dependence in Nanofluidic Osmotic Energy Conversion. Adv. Funct. Mater. 27 (2017) 1604302. http://doi.org/10.1002/adfm.201604302.

[14] R. B. Schoch, J. Y. Han, P. Renaud, Transport Phenomena in Nanofluidics. Rev. Mod. Phys. 80 (2008) 839-883. https://doi.org/10.1103/RevModPhys.80.839.

[15] I. Vlassiouk, S. Smirnov, Z. Siwy, Ionic Selectivity of Single Nanochannels. Nano Lett. 8 (2008) 1978-1985. http://dx.doi.org/10.1021/nl800949k.

[16] Z. Li, Y. Qiu, K. Li, J. Sha, T. Li, Y. Chen, Optimal Design of Graphene Nanopores for Seawater Desalination. J. Chem. Phys. 148 (2018) 014703. https://doi.org/10.1063/1.5002746.

[17] L. Cao, W. Guo, W. Ma, L. Wang, F. Xia, S. Wang, Y. Wang, L. Jiang, D. Zhu, Towards Understanding the Nanofluidic Reverse Electrodialysis System: Well Matched Charge Selectivity and Ionic Composition. Energy Environ. Sci. 4 (2011) 2259-2266. https://doi.org/10.1039/C1EE01088C.

[18] G. Laucirica, M. E. Toimil-Molares, C. Trautmann, W. A. Marmisollé, O. Azzaroni, Polyaniline for Improved Blue Energy Harvesting: Highly Rectifying Nanofluidic Diodes Operating in Hypersaline Conditions Via One-Step Functionalization. ACS Appl. Mater. Interfaces 12 (2020) 28148-28157. https://doi.org/10.1021/acsami.0c05102.

[19] G. Laucirica, A. G. Albesa, M. E. Toimil-Molares, C. Trautmann, W. A. Marmisollé, O. Azzaroni, Shape Matters: Enhanced Osmotic Energy Harvesting in Bullet-Shaped Nanochannels. Nano Energy 71 (2020) 104612. https://doi.org/10.1016/j.nanoen.2020.104612.

[20] A. Siria, P. Poncharal, A.-L. Biance, R. Fulcrand, X. Blase, S. T. Purcell, L. Bocquet, Giant Osmotic Energy Conversion Measured in a Single Transmembrane Boron Nitride Nanotube. Nature 494 (2013) 455-458. https://doi.org/10.1038/nature11876.

[21] J. Feng, M. Graf, K. Liu, D. Ovchinnikov, D. Dumcenco, M. Heiranian, V. Nandigana, N. R. Aluru, A. Kis, A. Radenovic, Single-Layer Mos2 Nanopores as Nanopower Generators. Nature 536 (2016) 197. http://doi.org/10.1038/nature18593.

[22] Y. Fu, X. Guo, Y. Wang, X. Wang, J. Xue, An Atomically-Thin Graphene Reverse Electrodialysis System for Efficient Energy Harvesting from Salinity Gradient. Nano Energy 57

(2019) 783-790. https://doi.org/10.1016/j.nanoen.2018.12.075.

[23] M. Macha, S. Marion, V. V. R. Nandigana, A. Radenovic, 2d Materials as an Emerging Platform for Nanopore-Based Power Generation. Nat. Rev. Chem. 4 (2019) 588-605. https://doi.org/10.1038/s41578-019-0126-z.

[24] W.-C. Huang, J.-P. Hsu, Ultrashort Nanopores of Large Radius Can Generate Anomalously High Salinity Gradient Power. Electrochim. Acta 353 (2020) 136613. https://doi.org/10.1016/j.electacta.2020.136613.

[25] S. Tseng, Y.-M. Li, C.-Y. Lin, J.-P. Hsu, Salinity Gradient Power: Optimization of Nanopore Size. Electrochim. Acta 219 (2016) 790-797. https://doi.org/10.1016/j.electacta.2016.10.014.

[26] Y. Lee, J. H. Kim, D.-K. Kim, Power Generation from Concentration Gradient by Reverse Electrodialysis in Anisotropic Nanoporous Anodic Aluminum Oxide Membranes. Energies 13 (2020) 904. https://doi.org/10.3390/en13040904.

[27] R. Long, Y. Zhao, Z. Kuang, Z. Liu, W. Liu, Hydrodynamic Slip Enhanced Nanofluidic Reverse Electrodialysis for Salinity Gradient Energy Harvesting. Desalination 477 (2020) 114263. https://doi.org/10.1016/j.desal.2019.114263.

[28] D. J. Rankin, D. M. Huang, The Effect of Hydrodynamic Slip on Membrane-Based Salinity-Gradient-Driven Energy Harvesting. Langmuir 32 (2016) 3420-3432. http://doi.org/10.1021/acs.langmuir.6b00433.

[29] B. D. Kang, H. J. Kim, M. G. Lee, D.-K. Kim, Numerical Study on Energy Harvesting from Concentration Gradient by Reverse Electrodialysis in Anodic Alumina Nanopores. Energy 86 (2015) 525-538. https://doi.org/10.1016/j.energy.2015.04.056.

[30] L. Cao, Q. Wen, Y. Feng, D. Ji, H. Li, N. Li, L. Jiang, W. Guo, On the Origin of Ion Selectivity in Ultrathin Nanopores: Insights for Membrane-Scale Osmotic Energy Conversion. Adv. Funct. Mater. 28 (2018) 1804189. https://doi.org/10.1002/adfm.201804189.

[31] M. Tagliazucchi, Y. Rabin, I. Szleifer, Transport Rectification in Nanopores with Outer Membranes Modified with Surface Charges and Polyelectrolytes. ACS Nano 7 (2013) 9085-9097. https://doi.org/10.1021/nn403686s.

[32] L. Ma, Z. Li, Z. Yuan, C. Huang, Z. S. Siwy, Y. Qiu, Modulation of Ionic Current Rectification in Ultrashort Conical Nanopores. Anal. Chem. 92 (2020) 16188-16196. https://doi.org/10.1021/acs.analchem.0c03989.

[33] Y. Qiu, Z. S. Siwy, M. Wanunu, Abnormal Ionic-Current Rectification Caused by Reversed Electroosmotic Flow under Viscosity Gradients across Thin Nanopores. Anal. Chem. 91 (2019) 996-1004. https://doi.org/10.1021/acs.analchem.8b04225.

[34] Y. Qiu, Optimal Voltage for Nanoparticle Detection with Thin Nanopores. Analyst 143 (2018) 4638-4645. http://dx.doi.org/10.1039/C8AN01270A.

[35] J. Chen, W. Xin, X.-Y. Kong, Y. Qian, X. Zhao, W. Chen, Y. Sun, Y. Wu, L. Jiang, L. Wen, Ultrathin and Robust Silk Fibroin Membrane for High-Performance Osmotic Energy Conversion. ACS Energy Lett. 5 (2020) 742-748. https://doi.org/10.1021/acsenergylett.9b02296.

[36] X. Huang, Z. Zhang, X.-Y. Kong, Y. Sun, C. Zhu, P. Liu, J. Pang, L. Jiang, L. Wen, Engineered Pes/Spes Nanochannel Membrane for Salinity Gradient Power Generation. Nano Energy 59 (2019) 354-362. https://doi.org/10.1016/j.nanoen.2019.02.056.

[37] X. Zhu, J. Hao, B. Bao, Y. Zhou, H. Zhang, J. Pang, Z. Jiang, L. Jiang, Unique Ion Rectification in Hypersaline Environment: A High-Performance and Sustainable Power

Generator System. Sci. Adv. 4 (2018) eaau1665. https://doi.org/10.1126/sciadv.aau1665.

[38] B. Tansel, J. Sager, T. Rector, J. Garland, R. F. Strayer, L. F. Levine, M. Roberts, M. Hummerick, J. Bauer, Significance of Hydrated Radius and Hydration Shells on Ionic Permeability During Nanofiltration in Dead End and Cross Flow Modes. Sep. Purif. Technol. 51 (2006) 40-47. https://doi.org/10.1016/j.seppur.2005.12.020.

[39] X. Liu, M. He, D. Calvani, H. Qi, K. B. S. S. Gupta, H. J. M. de Groot, G. J. A. Sevink, F. Buda, U. Kaiser, G. F. Schneider, Power Generation by Reverse Electrodialysis in a Single-Layer Nanoporous Membrane Made from Core–Rim Polycyclic Aromatic Hydrocarbons. Nat. Nanotechnol. 15 (2020) 307-312. https://doi.org/10.1038/s41565-020-0641-5.

[40] J. Gao, X. Liu, Y. Jiang, L. Ding, L. Jiang, W. Guo, Understanding the Giant Gap between Single-Pore- and Membrane-Based Nanofluidic Osmotic Power Generators. Small 15 (2019) 1804279. https://doi.org/10.1002/smll.201804279.

[41] E. L. Cussler, Diffusion: Mass Transfer in Fluid Systems, Cambridge University Press: Cambridge, 2009, pp 162.

[42] D.-K. Kim, C. Duan, Y.-F. Chen, A. Majumdar, Power Generation from Concentration Gradient by Reverse Electrodialysis in Ion-Selective Nanochannels. Microfluid. Nanofluidics 9 (2010) 1215-1224. https://doi.org/10.1007/s10404-010-0641-0.

[43] J. N. Israelachvili, Intermolecular and Surface Forces, 3rd ed., Academic Press: Burlington, MA, 2011.

[44] F. Xiao, D. Ji, H. Li, J. Tang, Y. Feng, L. Ding, L. Cao, N. Li, L. Jiang, W. Guo, A General Strategy to Simulate Osmotic Energy Conversion in Multi-Pore Nanofluidic Systems. Mater. Chem. Front. 2 (2018) 935-941. http://doi.org/10.1039/C8QM00031J.

[45] S. Marbach, L. Bocquet, Osmosis, from Molecular Insights to Large-Scale Applications. Chem. Soc. Rev 48 (2019) 3102-3144. http://doi.org/10.1039/C8CS00420J.

[46] R. C. Rollings, A. T. Kuan, J. A. Golovchenko, Ion Selectivity of Graphene Nanopores. Nat. Commun. 7 (2016) 11408. http://doi.org/10.1038/ncomms11408.